\documentstyle[aps,preprint]{revtex}
\begin{document}
\input{psfig}
\draft
\title{Analytical operator solution of master equations describing \\
phase-sensitive processes}
\author{A. Vidiella-Barranco\dag\footnote{E-mail: vidiella@ifi.unicamp.br}, 
Luis M. Ar\'{e}valo-Aguilar\ddag, 
and Hector Moya-Cessa\S\footnote{E-mail: hmmc@inaoep.mx}}
\address{\dag Instituto de F\'\i sica ``Gleb Wataghin'',
Universidade Estadual de Campinas, 13083-970  Campinas  SP  Brazil}  
\address{\ddag\S INAOE, Coordinaci\'on de Optica, Apdo. Postal 51 y 216, 72000 
Puebla, Pue., Mexico}
\date{\today}
\maketitle
\begin{abstract}
We present a method of solving master equations which may describe, in their 
most general form, phase sensitive processes such as decay and amplification.
We make use of the superoperator technique.
\end{abstract}
\pacs{42.50.Lc, 02.90.+p}
\newpage

\section{Introduction}

Master equations, or evolution equations for reduced density operators are 
of fundamental importance in the treatment of open systems in quantum theory. 
They are particularly relevant in the field of quantum optics \cite{Louis,ScMi}, 
where generally one has interest in studying the evolution of confined subsystems 
(a single mode of the electromagnetic field, for instance) ``coupled'' 
to external reservoirs. This may model a number of (phase-sensitive or not) processes, 
such as the decay of a single mode of the field confined in a lossy cavity \cite{Louis}, 
as well as amplification processes \cite{ScMi,zuba}. Interesting applications of 
phase-sensitive 
amplification, for example, are the reduction of noise in lasers through the injection 
of squeezed vacuum \cite{Gea}, as well as noise-free amplification via 
the two-photon correlated-emission-laser \cite{Scully}.

Nevertheless, complete operator solutions of master equations are not usually
presented. Only simple cases such as the decay of a single mode of the field into a
vacuum state reservoir ($T=0$) or stationary regime cases are the ones normally 
treated \cite{barnet,ScMi}. Instead, master equations are normally transformed into 
c-number equations either in the number or coherent state representations, for 
instance. It would be therefore interesting to obtain a direct solution of the master 
equation while in its operator form.

\section{The method}

In this paper we show how to solve exactly master equations of the form
\begin{equation}
\frac{\partial\hat{\rho}}{\partial t}=\left(\sum_{k=1}^4\gamma_k 
\hat{L}_k(\hat{a},\hat{a}^\dagger)\right)\hat{\rho},\label{eq:mastereq}
\end{equation}

where

\begin{eqnarray}
\hat{L}_1(\hat{a},\hat{a}^\dagger)\hat{\rho}&=&2\hat{a}\hat{\rho}\hat{a}^\dagger
-\hat{a}^\dagger\hat{a}\hat{\rho}-\hat{\rho}\hat{a}^\dagger\hat{a} \\
\hat{L}_2(\hat{a},\hat{a}^\dagger)\hat{\rho}&=&2\hat{a}^\dagger\hat{\rho}\hat{a}-
\hat{a}\hat{a}^\dagger\hat{\rho}-\hat{\rho}\hat{a}\hat{a}^\dagger \\
\hat{L}_3(\hat{a},\hat{a}^\dagger)\hat{\rho}&=&2\hat{a}\hat{\rho}\hat{a}-\hat{a}^2
\hat{\rho}-\hat{\rho}\hat{a}^2, 
\end{eqnarray}

and 
\begin{equation}
\hat{L}_4(\hat{a},\hat{a}^\dagger)\hat{\rho}=2\hat{a}^\dagger\hat{\rho}\hat{a}^\dagger-
\hat{a}^\dagger{}^2\hat{\rho}-\hat{\rho}\hat{a}^\dagger{}^2,
\end{equation}
being $\hat{a}$ and $\hat{a}^\dagger$ the bosonic annihilation and creation 
operators. The $\gamma$s are in general complex parameters which may represent gain and 
decay. For the density operator to be Hermitean, it is necessary that

\begin{equation}
\gamma_3=|\gamma_3|e^i\phi=\gamma_4^*.
\end{equation}

We may now apply the unitary transformation
\begin{equation}
\hat{\rho}=\hat{S}(\xi)\hat{\rho}'\hat{S}^\dagger(\xi)\label{eq:transfo}
\end{equation}

where $\hat{S}(\xi)$ is the squeeze operator
\begin{equation}
\hat{S}(\xi)=\exp\left(\frac{\xi^*\hat{a}^2}{2}-\frac{\xi\hat{a}^\dagger{}^2}{2}\right),
\label{eq:sqope}
\end{equation}
with $\xi=r\exp(i\varphi)$. If we now substitute Eq.(\ref{eq:transfo}) into
Eq.(\ref{eq:mastereq}), we obtain an equation for the transformed density operator 
$\hat{\rho}'$, or 
\begin{equation}
\frac{\partial\hat{\rho}'}{\partial t}=\left(\sum_{k=1}^4\gamma_k 
\hat{L}_k(\hat{b},\hat{b}^\dagger)\right)\hat{\rho}',\label{eq:mastereqt}
\end{equation}
where
\begin{equation}
\hat{b}=\hat{S}(\xi)^\dagger\hat{a}\hat{S}(\xi)=\mu\hat{a}+\nu\hat{a}^\dagger,\label{bdef}
\end{equation}
with
\begin{equation}
\mu=\cosh(r), \ \ \ \ \nu=-\sinh(r)\exp(i\varphi).
\end{equation}

We proceed in rewriting Eq.(\ref{eq:mastereqt}) in terms of the original creation and
annihilation operators as
\begin{eqnarray}
\frac{\partial\hat{\rho}'}{\partial t}=&&\hat{L}_1(\hat{a},\hat{a}^\dagger)\hat{\rho}'
\left(\gamma_1\mu^2+\gamma_2|\nu|^2+\mu\nu|\gamma_3|e^{i\phi}+\mu\nu^*|\gamma_3|
e^{-i\phi}\right)+ \nonumber \\
&&\hat{L}_2(\hat{a},\hat{a}^\dagger)\hat{\rho}'
\left(\gamma_1|\nu|^2+\gamma_2\mu^2+\mu\nu|\gamma_3|e^{i\phi}+\mu\nu^*|\gamma_3|
e^{-i\phi}\right)+ \nonumber \\
&&\hat{L}_3(\hat{a},\hat{a}^\dagger)\hat{\rho}'
\left(\gamma_1\mu\nu^*+\gamma_2\mu\nu^*+\mu^2|\gamma_3|e^{i\phi}+(\nu^*)^2|\gamma_3|
e^{-i\phi}\right)+ \nonumber \\
&&\hat{L}_4(\hat{a},\hat{a}^\dagger)\hat{\rho}'
\left(\gamma_1\mu\nu+\gamma_2\mu\nu+\nu^2|\gamma_3|e^{i\phi}+\mu^2|\gamma_3|
e^{-i\phi}\right).\label{eq:eqtrans}
\end{eqnarray}

We note that one can choose $\phi$ and $r$ in such a way that the last two terms in
Eq.(\ref{eq:eqtrans}) become zero. The appropriate choices for these parameters are
\begin{equation}
\varphi=-\phi,
\end{equation}
and 
\begin{equation}
\mbox{tanh}(2r)=\frac{2|\gamma_3|}{\gamma_1+\gamma_2}.
\end{equation}

With the above choice of parameters, then we can write Eq.(\ref{eq:eqtrans}) as
\begin{equation}
\frac{\partial\hat{\rho}'}{\partial t}=\left[\tilde{\gamma}_1 
\hat{L}_1(\hat{a},\hat{a}^\dagger)+\tilde{\gamma}_2\hat{L}_2(\hat{a},\hat{a}^\dagger)
\right]\hat{\rho}',\label{eq:mmaeq}
\end{equation}

where
\begin{equation}
\tilde{\gamma}_1=\gamma_1\mu^2+\gamma_2|\nu|^2+\mu\nu|\gamma_3|e^{i\phi}+
\mu\nu^*|\gamma_3|e^{-i\phi},
\end{equation}

and
\begin{equation}
\tilde{\gamma}_2=\gamma_1|\nu|^2+\gamma_2\mu^2+\mu\nu|\gamma_3|e^{i\phi}+
\mu\nu^*|\gamma_3|e^{-i\phi}.
\end{equation}

It follows from the relations above that

\begin{equation}
\tilde{\gamma}_1-\tilde{\gamma}_2=\gamma_1-\gamma_2,
\end{equation}
and
\begin{equation}
\tilde{\gamma}_1+\tilde{\gamma}_2=(\gamma_{1} + \gamma_{2})\,\mbox{sech}(2r).
\end{equation}

We now write the new master equation (\ref{eq:mmaeq}) in a more 
convenient form \cite{Aguilar}
\begin{equation}
\frac{\partial\hat{\rho}'}{\partial t}=\left(\hat{J}_1+\hat{J}_2+\hat{J}_3-2\tilde{\gamma}_2
\right)\hat{\rho}',\label{eq:smaeq}
\end{equation}

where we have defined the following super-operators
\begin{equation}
\hat{J}_1\hat{\rho}'=2\tilde{\gamma}_1\hat{a}\hat{\rho}'\hat{a}^\dagger,
\end{equation}

\begin{equation}
\hat{J}_2\hat{\rho}'=2\tilde{\gamma}_2\hat{a}^\dagger\hat{\rho}'\hat{a},
\end{equation}
and
\begin{equation}
\hat{J}_3\hat{\rho}'=-(\tilde{\gamma}_1+\tilde{\gamma}_2)(\hat{a}^\dagger\hat{a}
\hat{\rho}'+
\hat{\rho}'\hat{a}^\dagger\hat{a}),
\end{equation}

From the equations above, we can write the formal solution of Eq.(\ref{eq:smaeq}) as
\begin{equation}
\hat{\rho}'(t)=\exp(-2\tilde{\gamma}_2 t)\exp\left[\left(\hat{J}_1+\hat{J}_2+\hat{J}_3
\right)t
\right]\hat{\rho}'(0).\label{eq:forsol}
\end{equation}

It is not difficult to show that the superoperators $\hat{J}_1$, $\hat{J}_2$ and 
$\hat{J}_3$  obey the following commutation
relations \cite{remar}
\begin{equation}
\left[\hat{J}_2,\hat{J}_1\right]\hat{\rho}'=\left[\frac{4\tilde{\gamma}_1
\tilde{\gamma}_2}{\tilde{\gamma}_1+
\tilde{\gamma}_2}\hat{J}_3-4\tilde{\gamma}_1\tilde{\gamma}_2\right]\hat{\rho}',
\end{equation}
\begin{equation}
\left[\hat{J}_1,\hat{J}_3\right]\hat{\rho}'=-2\left(\tilde{\gamma}_1+\tilde{\gamma}_2
\right)
\hat{J}_1\hat{\rho}',
\end{equation}
and
\begin{equation}
\left[\hat{J}_2,\hat{J}_3\right]\hat{\rho}'=2\left(\tilde{\gamma}_1+\tilde{\gamma}_2
\right)
\hat{J}_2\hat{\rho}'.
\end{equation}

In order to disentangle the exponential in Eq.(\ref{eq:forsol}), we propose the 
ansatz
\begin{equation}
\hat{\rho}'(t)=\exp(-2\tilde{\gamma}_2 t)\exp[f_3(t)]\exp[f_2(t)\hat{J}_2]
\exp[f_0(t)\hat{J}_3]\exp[f_1(t)\hat{J}_1(t)]\hat{\rho}'(0)\label{eq:ansat}
\end{equation}

By inserting $\hat{\rho}'(t)$ in Eq.(\ref{eq:ansat}) above into equation in  
Eq.(\ref{eq:smaeq}), we obtain the following system of differential equations
for the functions $f_i$:

\begin{equation}
\frac{df_0}{dt}+\frac{4\tilde{\gamma}_1\tilde{\gamma}_2}{\tilde{\gamma}_1+
\tilde{\gamma}_2}f_2=1,\label{eq:eqfo}
\end{equation}
\begin{equation}
\frac{df_1}{dt}\exp[2(\tilde{\gamma}_1+\tilde{\gamma}_2)]=1,
\end{equation}
\begin{equation}
\frac{df_2}{dt}+2\frac{df_0}{dt}f_2+(\tilde{\gamma}_1+\tilde{\gamma}_2)+
4\tilde{\gamma}_1\tilde{\gamma}_2f^2_2=1,
\end{equation}
and
\begin{equation}
\frac{df_3}{dt}-4\tilde{\gamma}_1\tilde{\gamma}_2f_2=0.\label{eq:eqf3}
\end{equation}

Although the system is a non-linear one, its solution is
rather straightforward. In order to have the condition 
$\hat{\rho}'(t=0)=\hat{\rho}'(0)$ satisfied, we should have $f_i(0)=0$ ($i=0,1,2,3$)
as initial conditions for the set of Eqs.(\ref{eq:eqfo})-(\ref{eq:eqf3}). 
The result is

\begin{equation}
f_0=\frac{\kappa}{\tilde{\gamma}_1+\tilde{\gamma}_2}t+
\frac{1}{\tilde{\gamma}_1+\tilde{\gamma}_2}
\ln\left(\frac{\kappa + \Gamma (t)}{\kappa}\right),
\label{eq:fzero}
\end{equation}
\begin{equation}
f_1=f_2=\frac{1}{2}\frac{\Gamma (t)}{\kappa+\Gamma(t)},
\label{eq:fumdois}
\end{equation}
and
\begin{equation}
f_3=2\tilde{\gamma}_2t-\ln\left(\frac{\kappa + \Gamma (t)}{\kappa}\right),
\label{eq:ftres}
\end{equation}
where $\Gamma (t)=(1-\exp[-2\kappa t])\tilde{\gamma}_2$.

By finally defining the dimensionless superoperators
\begin{equation}
\hat{L}_-\hat{\rho}\doteq \hat{\underline{b}}\hat{\rho}\hat{\underline{b}}^\dagger, \ \ \ \
\hat{L}_+\hat{\rho}\doteq \hat{\underline{b}}^\dagger \hat{\rho} \hat{\underline{b}},
\ \ \ \and \ \ \ 
\hat{L}_3\hat{\rho}\doteq \hat{\underline{b}}^\dagger\hat{\underline{b}}\hat{\rho}+
\hat{\rho}\hat{\underline{b}}^\dagger\hat{\underline{b}}+\hat{\rho},
\label{btilde}
\end{equation}
where the superoperators $\hat{L}_-$, $\hat{L}_+$ and $\hat{L}_3$ obey the commutation 
relations $[\hat{L}_-,\hat{L}_+]\hat{\rho}=\hat{L}_3\hat{\rho}$ and
$[\hat{L}_3,\hat{L}_{\pm}]\hat{\rho}=\pm 2\hat{L}_{\pm}\hat{\rho}$,
we can write the solution of (\ref{eq:mastereq}) in the form, 
\begin{equation}
\hat{\rho}(t) = e^{\kappa t}
e^{\frac{\Gamma (t)}{\kappa+\Gamma(t)}\hat{L}_+}
\left[\frac{\kappa e^{-\kappa t}}{\kappa + \Gamma (t)}\right]^{\hat{L}_3}
e^{\frac{\tilde{\gamma}_1}{\tilde{\gamma}_2}
\frac{\Gamma (t)}{\kappa+\Gamma(t)}\hat{L}_-}
\label{finalrho}
\hat{\rho}(0),
\end{equation}
where in (\ref{btilde}) we have defined
\begin{equation}
\hat{\underline{b}}=\hat{S}(\xi)\hat{a}\hat{S}^\dagger(\xi)=\mu\hat{a}-\nu\hat{a}^\dagger,
\end{equation}
which should be compared with Eq. (\ref{bdef}). Note that if in Eq. (\ref{finalrho})
we set the parameters $\xi=0$ and $\tilde{\gamma}_2=0$ we recover the usual 
solution for a dissipative cavity at zero temperature as well as for the phase 
insensitive case.

We have therefore obtained the full solution of the master equation for $\hat{\rho}'$ 
[see Eq.(\ref{eq:smaeq})], given by (\ref{finalrho}). We remind that we still have to apply the
transformation in Eq.(\ref{eq:transfo}) in order to recover the original density operator 
$\hat{\rho}$. 

It would be appropriate now to show which interesting cases could be easily treated 
by employing our solution. For instance, if we make $\gamma_1=\gamma\,(\overline{n}+1)/2$, 
$\gamma_2=\gamma\,\overline{n}/2$, $\gamma_3=-\gamma\,M^*/2$ and $\gamma_4=-\gamma\,M/2$, we 
will obtain the standard equation describing the decay of a bosonic field into a phase-sensitive
reservoir \cite{ScMi}. If on the other hand $\gamma_1={\cal{A}}\,\overline{n}/2$ and 
$\gamma_2={\cal{A}}\,(\overline{n}+1)/2$, but keeping $\gamma_3=-{\cal{A}}\,M^*/2$ and 
$\gamma_4=-{\cal{A}}\,M/2$, we will have the standard equation describing 
phase-sensitive amplification \cite{ScMi}. 
The constants $\gamma$ and ${\cal{A}}$ represent decay and gain, respectively, and
$M$ ($\overline{n}$) are connected with phase-sensitive (or not) reservoir fluctuations. 
Each of this two cases are relevant in the problems of field decay and in the 
reduction of noise in lasers \cite{Gea}. We remark that our operator 
solution turns unnecessary the convertion of the master equation into sometimes 
cumbersome c-number equations. 

Let us finally consider as an example an initial density matrix of the form
\begin{equation}
\hat{\rho}(0)=\hat{S}(\xi)|0\rangle \langle 0|\hat{S}^\dagger(\xi),
\end{equation}
or a squeezed vacuum state.
After inserting it into (\ref{finalrho}) it is
easy to show that the density operator of the field at a time $t$ may be written as
\begin{equation}
\hat{\rho}(t)=\frac{\kappa}{\kappa+\Gamma(t)}
\sum_{m=0}^\infty \left[ \frac{\Gamma(t)}{\kappa+\Gamma(t)} \right ]^m
|m,\xi\rangle \langle m,\xi|,
\end{equation}
where $|m,\xi\rangle = \hat{S}(\xi)|m\rangle$ are the squeezed number states
\cite{satia,deoliveira}.
We may now to follow the state's evolution in phase space,
using for instance the $Q$-function \cite{wigner}, defined as
\begin{equation}
Q(t)=\frac{1}{\pi}\langle\beta|\hat{\rho}(t)|\beta\rangle,
\end{equation}
 
where $|\beta\rangle$ is a coherent state.
In our case the $Q$-function is given by
\begin{equation}
Q(t)=\frac{1}{\pi}\frac{\kappa}{\kappa+\Gamma(t)}\sum_{m=0}^\infty
\left[\frac{\Gamma(t)}{\kappa+\Gamma(t)}\right]^m|\langle\beta|m,\xi\rangle|^2,
\label{qfuncomp}
\end{equation}
with
\begin{equation}
\langle\beta|m,\xi\rangle=\exp(-|\beta|^2/2)\sum_{n=0}^\infty  G_{nm} (-\beta^*)^n,
\end{equation}
and where \cite{satia}
\begin{equation}
G_{nm}=\left\{ \begin{array}{ll}
(-1)^{n+m/2}\left(\frac{m!}{\cosh r}\right)^{1/2}\left(\frac{\tanh r}{2}\right)
^{n+m/2}\sum_{l=0} \frac{\left(\frac{-4}{\sinh^2 r}\right)^l}{(2l)!
\left(\frac{n}{2}-l\right)!\left(\frac{m}{2}-l\right)!}& \mbox{if $n,m$ even} \\
(-1)^{n+m/2-3/2}\left(\frac{m!}{\cosh^3 r}\right)^{1/2}\left(\frac{\tanh r}{2}\right)
^{(n+m)/2-1}
\sum_{l=0} \frac{\left(\frac{-4}{\sinh^2 r}\right)^l}{(2l+1)!
\left(\frac{n-1}{2}-l\right)!\left(\frac{m-1}{2}-l\right)!}& \mbox{if $n,m$ odd} \\
0 & \mbox{otherwise}
\end{array}
\right.
\end{equation}

In figure 1 it is illustrated the $Q$-function in (\ref{qfuncomp}) for 
different times. At $t=0$ we have the $Q$ function of the initial squeezed
vacuum state. As time goes on, thermal fluctuations of the reservoir cause a
``spread'' of the $Q$ function, associated to the increase in the quadrature
noise. Nevertheless, due to the phase-sensitive properties of the reservoir, 
the noise assymmetry characteristic of squeezed states is somehow preserved, 
as it may be seen in figure 1.

\section{Conclusions}

To summarize, we have presented an alternative way of treating problems involving certain 
types of master equations. Our operator solution may be useful to retrieve any sort of 
information related to the evolution of sub-systems having 
specific initial conditions. This contrasts with the usual approaches, where  
only partial information, such as mean values of amplitudes and/or diffusion 
coefficients is normally obtained.

{\it Note added in proof:} While preparing the answer to the referee, we became
aware 
of the paper by Dung and Kn\"oll \cite{Dung} where they also solve Eq. (\ref{eq:mastereq})
but using Fokker-Planck equations instead. We would like to stress that our method of
solving the master equation (\ref{eq:mastereq}) is considerably more straightforward and 
simple.

\acknowledgments

The authors would like to thank the Mexican Consejo Nacional de Ciencia 
y Tecnolog\'\i a (CONACyT) and the Brazilian Conselho Nacional de Desenvolvimento
Cient\'\i fico e Tecnol\'ogico (CNPq) for support.

\begin{figure}[hp]
\vspace{0.1cm}
\centerline{\hspace{1.0cm}\psfig{figure=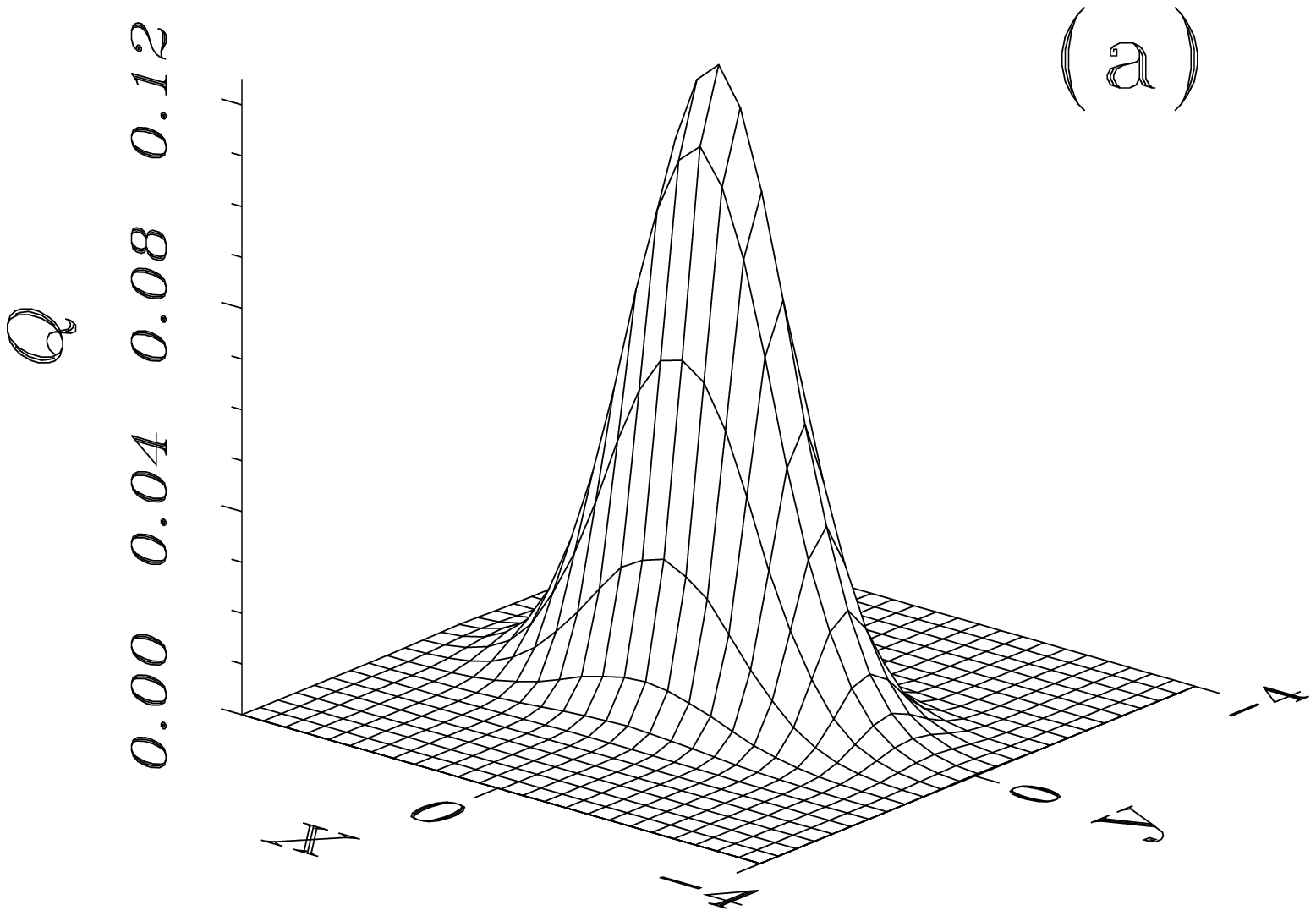,height=5cm,width=9cm}}
\end{figure}
\vspace{1cm}
\begin{figure}[hp]
\vspace{0.1cm}
\centerline{\hspace{1.0cm}\psfig{figure=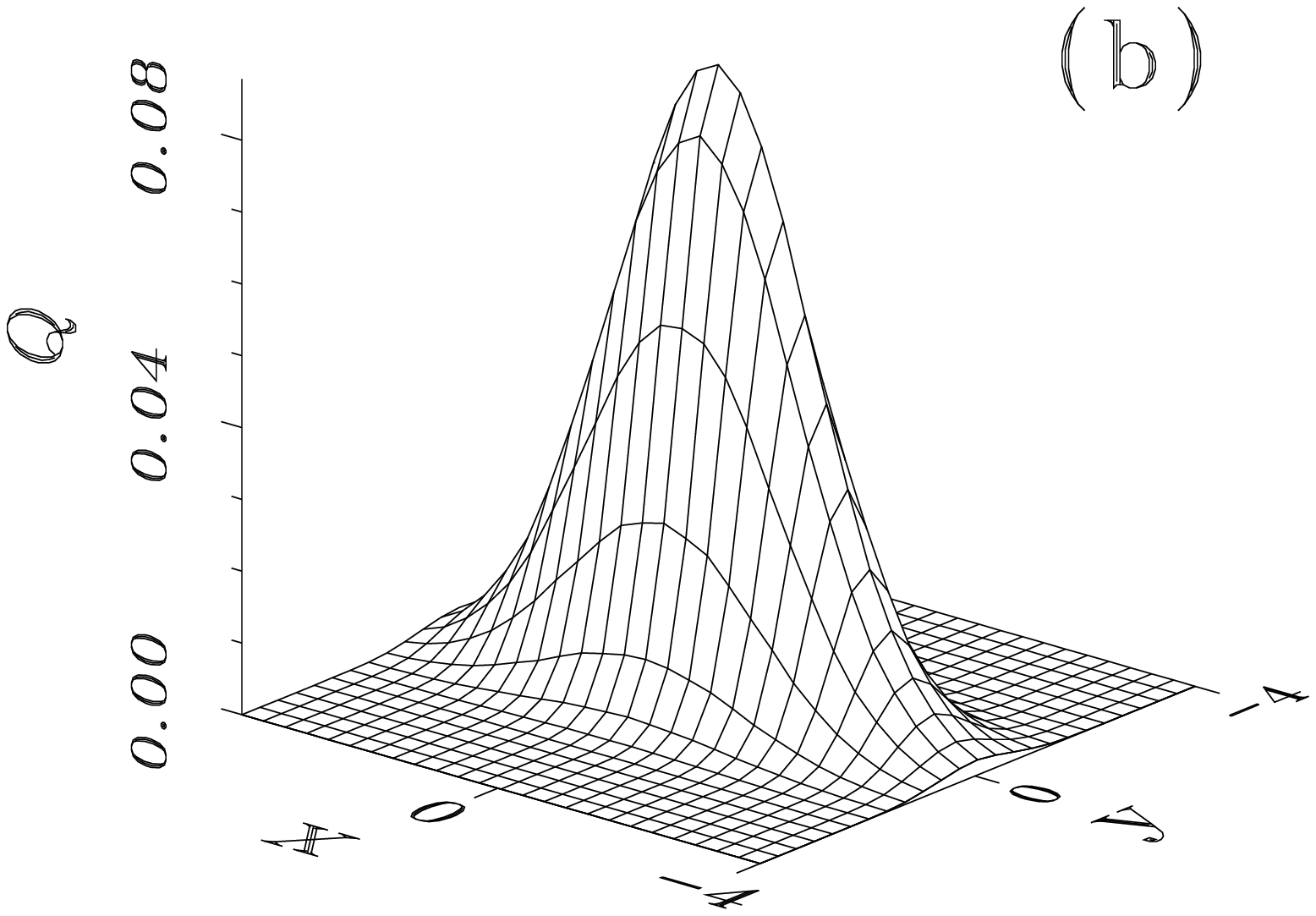,height=5cm,width=9cm}}
\end{figure}
\vspace{1cm}
\begin{figure}[hp]
\vspace{0.1cm}
\centerline{\hspace{1.0cm}\psfig{figure=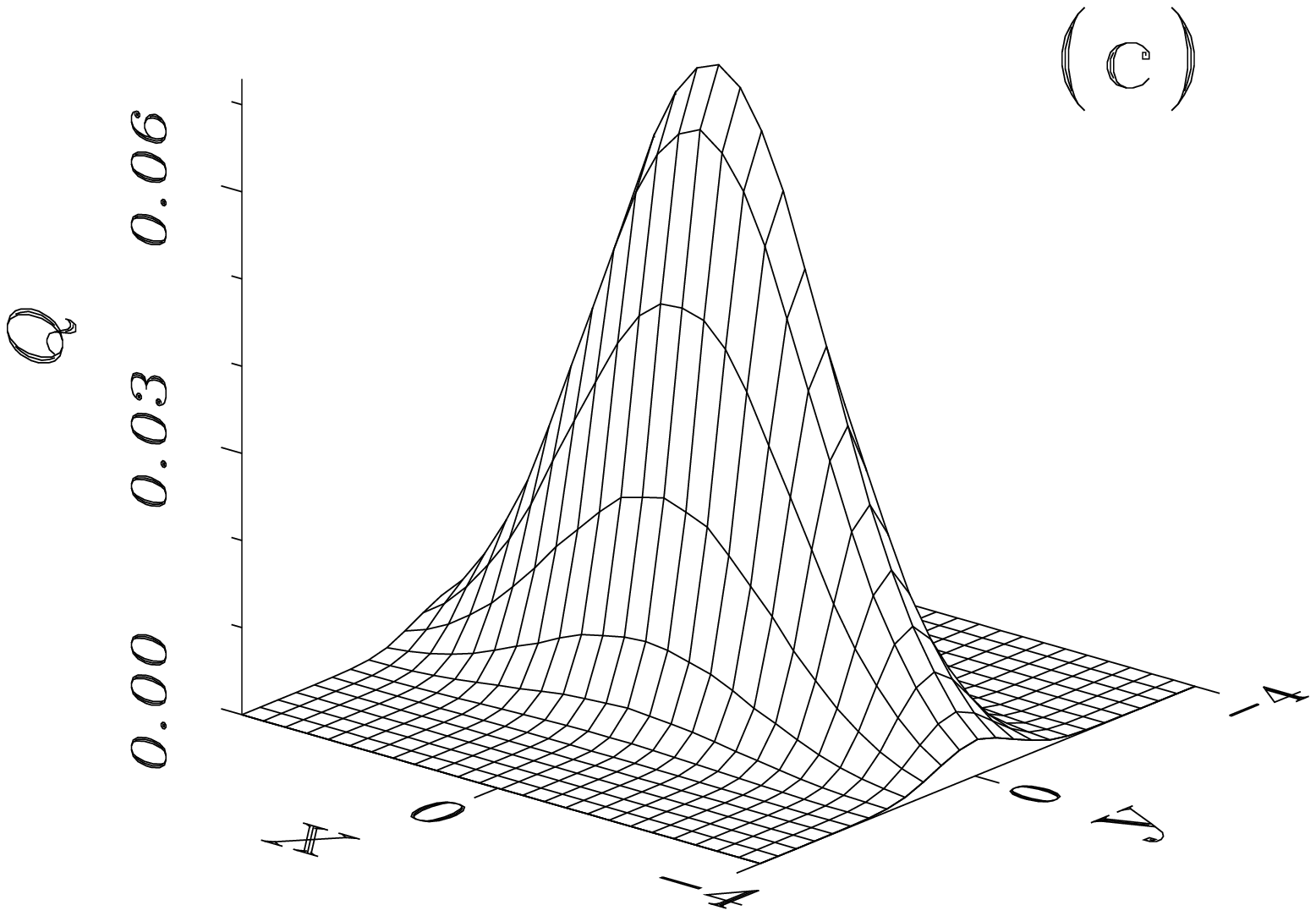,height=5cm,width=9cm}}
\vspace{1.5cm}
\caption{$Q$ function of the cavity field initially in a squeezed vacuum state.
a) at $\kappa t=0$, b) at $\kappa t=0.5$, and c) at $\kappa t=1.0$. It has been taken
$\xi=r=0.7$ real, and $\tilde{\gamma}_2=1$.}
\end{figure}


\begin{references}
\bibitem{remar}
Note that the commutators $[\hat{J}_i,\hat{J}_j]$ resemble those belonging to the 
$SU(2)$ group.
\end{references}
\end{document}